\documentclass[amsmath,amssymb,superscriptaddress,nobalancelastpage,nature,twocolumn]{revtex4-1}

\usepackage{graphicx}
\usepackage{varioref}
\usepackage{xr-hyper}
\usepackage{xcolor}
\usepackage{nicefrac}
\usepackage{xfrac}
\usepackage{hyperref}
\hypersetup{colorlinks,linkcolor=blue,urlcolor=blue,citecolor=blue}
\usepackage[normalem]{ulem}
\usepackage{lineno}
\usepackage{amsmath} 
\usepackage{amssymb}
\usepackage{physics}
\usepackage[up,bf,tiny,raggedright,compact]{titlesec}

\begin{document}

\title{Magnetotransport of dirty-limit van Hove singularity quasiparticles}

\author{Yang Xu}
\affiliation{Physik-Institut, Universit\"{a}t Z\"{u}rich, Winterthurerstrasse 190, CH-8057 Z\"{u}rich, Switzerland}

\author{Frantisek Herman}
\affiliation{Department of Experimental Physics, Comenius University, Mlynsk\'a Dolina F2, 842 48 Bratislava, Slovakia}
\affiliation{Institute for Theoretical Physics, ETH Z\"{u}rich, H\"{o}nggerberg, CH-8093 Z\"{u}rich, Switzerland}

\author{Veronica Granata}
\affiliation{CNR-SPIN, I-84084 Fisciano, Salerno, Italy}
\affiliation{Dipartimento di Fisica "E.R.~Caianiello", Universit\`{a} di Salerno, I-84084 Fisciano, Salerno, Italy}

\author{Daniel Destraz}
\affiliation{Physik-Institut, Universit\"{a}t Z\"{u}rich, Winterthurerstrasse 190, CH-8057 Z\"{u}rich, Switzerland}

\author{Lakshmi Das}
\affiliation{Physik-Institut, Universit\"{a}t Z\"{u}rich, Winterthurerstrasse 190, CH-8057 Z\"{u}rich, Switzerland}

\author{Jakub Vonka}
\affiliation{Laboratory for Micro and Nanotechnology, Paul Scherrer Institut, Forschungsstrasse 111, CH-5232 Villigen PSI, Switzerland}
\affiliation{Laboratory for Neutron and Muon Instrumentation, Paul Scherrer Institut, Forschungsstrasse 111, CH-5232 Villigen PSI, Switzerland}

\author{Simon Gerber}
\affiliation{Laboratory for Micro and Nanotechnology, Paul Scherrer Institut, Forschungsstrasse 111, CH-5232 Villigen PSI, Switzerland}

\author{Jonathan Spring}
\affiliation{Physik-Institut, Universit\"{a}t Z\"{u}rich, Winterthurerstrasse 190, CH-8057 Z\"{u}rich, Switzerland}

\author{Marta Gibert}
\affiliation{Physik-Institut, Universit\"{a}t Z\"{u}rich, Winterthurerstrasse 190, CH-8057 Z\"{u}rich, Switzerland}

\author{Andreas Schilling}
\affiliation{Physik-Institut, Universit\"{a}t Z\"{u}rich, Winterthurerstrasse 190, CH-8057 Z\"{u}rich, Switzerland}

\author{Xiaofu Zhang}
\affiliation{Physik-Institut, Universit\"{a}t Z\"{u}rich, Winterthurerstrasse 190, CH-8057 Z\"{u}rich, Switzerland}

\author{Shiyan Li}
\affiliation{State Key Laboratory of Surface Physics, Department of Physics, and Laboratory of Advanced Materials, Fudan University, Shanghai 200438, China}

 \author{Rosalba Fittipaldi}
\affiliation{CNR-SPIN, I-84084 Fisciano, Salerno, Italy}
\affiliation{Dipartimento di Fisica "E.R.~Caianiello", Universit\`{a} di Salerno, I-84084 Fisciano, Salerno, Italy}

\author{Mark H. Fischer}
\affiliation{Physik-Institut, Universit\"{a}t Z\"{u}rich, Winterthurerstrasse 190, CH-8057 Z\"{u}rich, Switzerland}

\author{Antonio Vecchione}
\affiliation{CNR-SPIN, I-84084 Fisciano, Salerno, Italy}
\affiliation{Dipartimento di Fisica "E.R.~Caianiello", Universit\`{a} di Salerno, I-84084 Fisciano, Salerno, Italy}

\author{Johan Chang}
\email{johan.chang@physik.uzh.ch}
\affiliation{Physik-Institut, Universit\"{a}t Z\"{u}rich, Winterthurerstrasse 190, CH-8057 Z\"{u}rich, Switzerland}

\date{\today}




\maketitle

\section*{Abstract}
\noindent \textbf{Tuning of electronic density-of-states singularities is a common route to unconventional metal physics. Conceptually, van Hove singularities are realized only in clean two-dimensional systems. Little attention has therefore been given to the disordered (dirty) limit. Here, we provide a magnetotransport study of the dirty metamagnetic system calcium-doped strontium ruthenate. Fermi liquid properties persist across the metamagnetic transition, but with an unusually strong variation of the Kadowaki-Woods ratio. This is revealed by a strong decoupling of inelastic electron scattering and electronic mass inferred from density-of-state probes. We discuss this Fermi liquid behavior in terms of a magnetic field tunable van Hove singularity in the presence of disorder. More generally, we show how dimensionality and disorder control the fate of transport properties across metamagnetic transitions.} \\[2mm]

\section*{Introduction}
\noindent In two-dimensional systems, saddle points in the electronic band structure generate a diverging density of states (DOS), a so-called van Hove singularity
 (VHS)~\cite{volovik_topological_2017}.
A divergent DOS at the Fermi level renders a system susceptible to instabilities like charge/spin density wave order or unconventional superconductivity. 
Gate-tuned superconductivity in magic angle bilayer graphene has, for example, been proposed to be connected to VHS physics~\cite{cao_unconventional_2018,yuan_magic_2019}.
A VHS is also found in high-temperature cuprate superconductors, and recently it has been associated with the onset of the mysterious pseudogap phase~\cite{wu_pseudogap_2018,doiron-leyraud_pseudogap_2017}. It is debated whether the surrounding non-Fermi liquid behavior is originating from a quantum criticality or a VHS scenario~\cite{buhmann_numerical_2013}. In the ruthenates, the VHS governs many interesting electronic properties. For example, the VHS can be tuned to the Fermi level by application of a magnetic field~\cite{GrigeraScience2001,grigera_disorder-sensitive_2004,BorziScience2007,LesterNatMat2015} or uniaxial~\cite{BarberPRL2018_full}, biaxial~\cite{BurganovPRL2016_full} and chemical pressure~\cite{kikugawa_band-selective_2004,ShenPRL2007}. In Sr$_3$Ru$_2$O$_7$, a magnetic field of 8 T along the $c$ axis triggers a spin density phase around which non-Fermi liquid transport behavior is observed~\cite{GrigeraScience2001,grigera_disorder-sensitive_2004,BorziScience2007,LesterNatMat2015,TokiwaPRL2016}. Similar non-Fermi liquid behavior is found in (Sr, Ba)$_2$RuO$_4$ upon application of pressure or strain~\cite{BarberPRL2018_full,BurganovPRL2016_full}. Finally, metamagnetic transitions in systems such as Sr$_3$Ru$_2$O$_7$, CeRu$_2$Si$_2$ and Ca$_{1.8}$Sr$_{0.2}$RuO$_4$ have been assigned to DOS anomalies near the Fermi level~\cite{tamai_fermi_2008,DaouPRL2006,NakatsujiPRL2003}. 

 \begin{figure*}
\begin{center}
 		\includegraphics[width=0.85\textwidth]{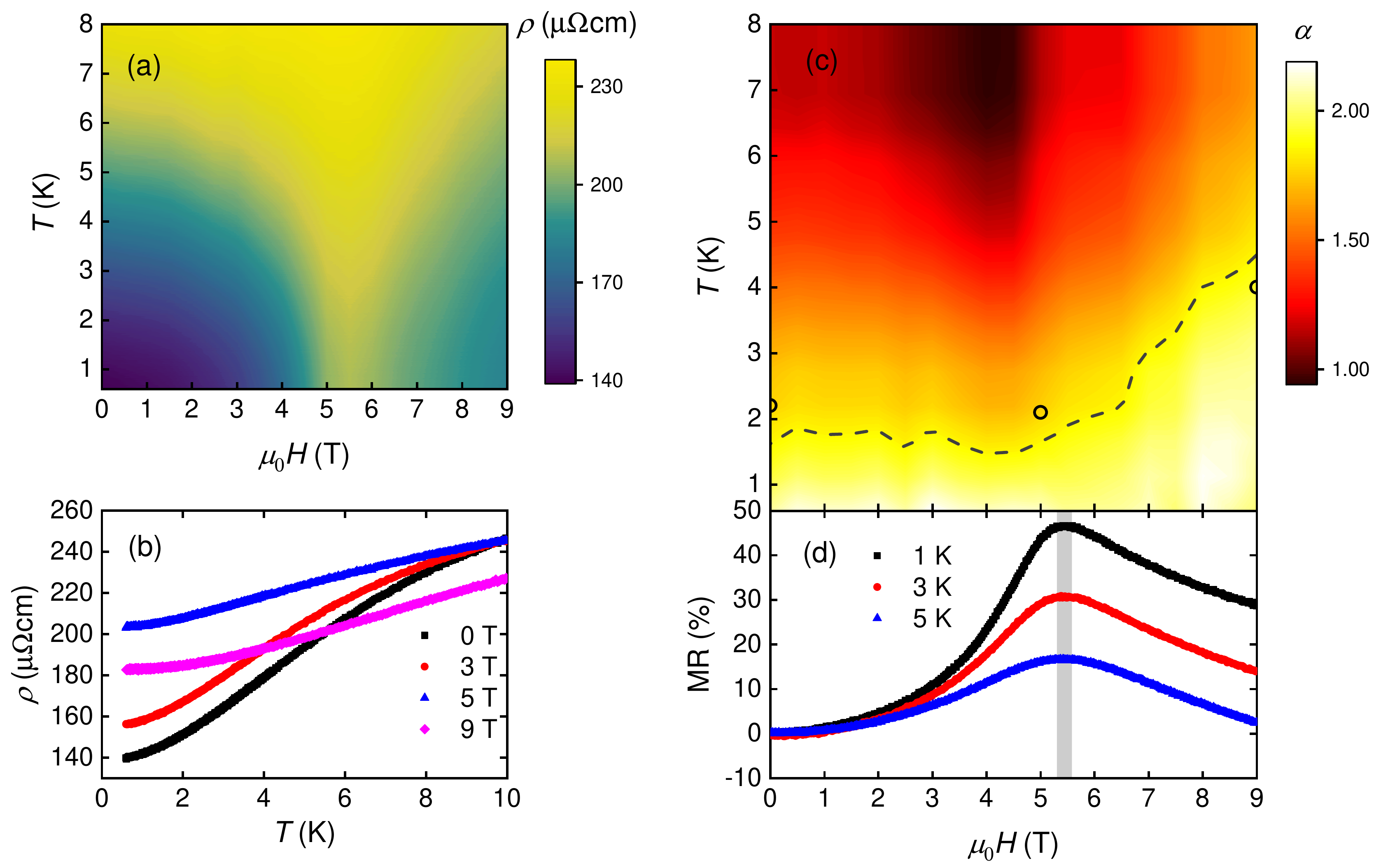}
 	\end{center}
\centering
\caption{\textbf{Magnetoresistance across the metamagnetic transition in Ca$_{1.8}$Sr$_{0.2}$RuO$_4$.} (a) The resistivity $\rho$ of Ca$_{1.8}$Sr$_{0.2}$RuO$_4$ as a function of temperature $T$ and magnetic field $H$. (b) The temperature dependence of $\rho$ for selected fields. (c) The exponent $\alpha$ in the $H$-$T$ space with the resistivity of Ca$_{1.8}$Sr$_{0.2}$RuO$_4$ fitted to $\rho$ = $\rho_0$ + $CT^{\alpha}$, with $C$ being the coefficient. The Fermi liquid cutoff temperatures $T${$\rm_{FL}$} at different magnetic fields are superimposed: the dashed line represents the contour line as the boundary between $\alpha~\sim~2$ and $\alpha~<~2$, while the open symbols represent $T${$\rm_{FL}$} as extracted from Fig. 2(a) (the error bars are smaller than the symbols). (d)  Magnetoresistance [$\frac{\rho (H) - \rho(0~\rm{T})}{\rho(0~\rm{T})}$] isotherms for selected temperatures. The gray shaded area indicates the maximum around $H_m$.}
\end{figure*}

Despite the expected connection between an ideal VHS and unconventional electronic properties observed in a wide range of materials, the effect of disorder and dimensionality has received little attention.  Quasiparticles in layered materials are neither constrained perfectly in two dimensions nor are their lifetime infinite. Both effects, dimensionality and disorder or electron correlations, broaden the DOS anomaly~\cite{horio_three-dimensional_2018} and hence, potentially change the ideal VHS physics substantially. 

Here, we address electronic transport properties of a quasi-two-dimensional disordered system for which 
the VHS is aligned with the Fermi level by an external magnetic field. Magnetotransport anomalies in Ca$_{1.8}$Sr$_{0.2}$RuO$_4$ are directly linked to the metamagnetic transition. Although Fermi liquid properties are preserved across the metamagnetic transition, the electronic scattering processes are highly unusual. In particular, we report a decoupling of the inelastic electron scattering from the electronic mass. This results in a five-fold variation of the Kadowaki-Woods ratio across the metamagnetic transition. Our observations are presented in a broader context of Fermi liquid/non-Fermi liquid properties across metamagnetic transitions in strongly correlated electron systems with DOS anomalies. Specifically, the role of dimensionality and disorder in the context of VHS physics is discussed along with possible multiband scenarios for the strong variation of the Kadowaki-Woods ratio. \\[2mm]

\section*{Results}
\noindent \textbf{Magnetotransport:} The temperature dependence of the resistivity $\rho$ measured on Ca$_{1.8}$Sr$_{0.2}$RuO$_4$ at various magnetic fields, is shown in Figs. 1(a) and 1(b). A region of enhanced resistivity fans out around the metamagnetic transition at $\mu_0 H_m \sim$ 5.5~T (see Supplementary Note 1 and Supplementary Figure 1) in the $\rho(H,T)$ plot [Fig. 1(a)]. Insights into the scattering mechanisms are commonly gained by analyzing $\rho$ = $\rho_0$ + $CT^{\alpha}$ with $C$ being a constant. The temperature-independent term $\rho_0$ is allowed to vary with field. Figure 1(c) shows the $H-T$ plot of the exponent $\alpha$ for Ca$_{1.8}$Sr$_{0.2}$RuO$_4$ obtained from this procedure. The low-temperature yellow region demonstrates that Fermi liquid behavior ($\alpha$ $\sim$ 2) is found at all fields across $H_m$. The Fermi liquid cutoff temperature $T${$\rm_{FL}$} remains constant below $H_m$ and increases above the transition. Magnetoresistance (MR) isotherms, defined by [$\frac{\rho (H) - \rho(0~\rm{T})}{\rho(0~\rm{T})}$], 
all exhibit a maximum around $H_m$ that broadens with increasing $T$ [Fig. 1(d)].

\noindent \textbf{Fermi liquid analysis:} Since Fermi liquid behavior is observed at low temperature for all fields, we fix $\alpha=2$ and fit with $\rho$ = $\rho_0$ + $AT^2$ [see Fig.~2(a)], where $A$ is the inelastic electron-electron scattering coefficient. In addition to the Fermi liquid cutoff temperature $T${$\rm_{FL}$} indicated by arrows in Fig. 2(a), we identify another temperature scale $T${$\rm_{SM}$}, above which a strange metal behavior $\rho \sim T$ is observed for all fields, as shown in Fig. 2(b). The resulting $\rho_0$ and $A$ from the analysis in Fig. 2(a) are plotted versus magnetic field in Figs. 2(c) and 2(d), respectively. The Kadowaki-Woods ratio (KWR) $A/\gamma^2$ ($\gamma$ being the Sommerfeld coefficient) is plotted in Fig. 2(e). We stress that our higher value of $\rho_0$ compared to Ref.~\cite{BalicasPRL2005} is not due to a lower quality of our sample (see Supplementary Note 2). While the field dependence of $\rho_0$ closely tracks the MR isotherms, $A$ decreases by a factor of three across $H_m$. Two key observations are revealed by our magnetotransport experiment: Across the metamagnetic transition, (1) the Fermi liquid state persists at low temperatures and (2) the inelastic scattering coefficient $A$ undergoes a dramatic drop.  

\begin{figure*}
\begin{center}
 		\includegraphics[width=0.9\textwidth]{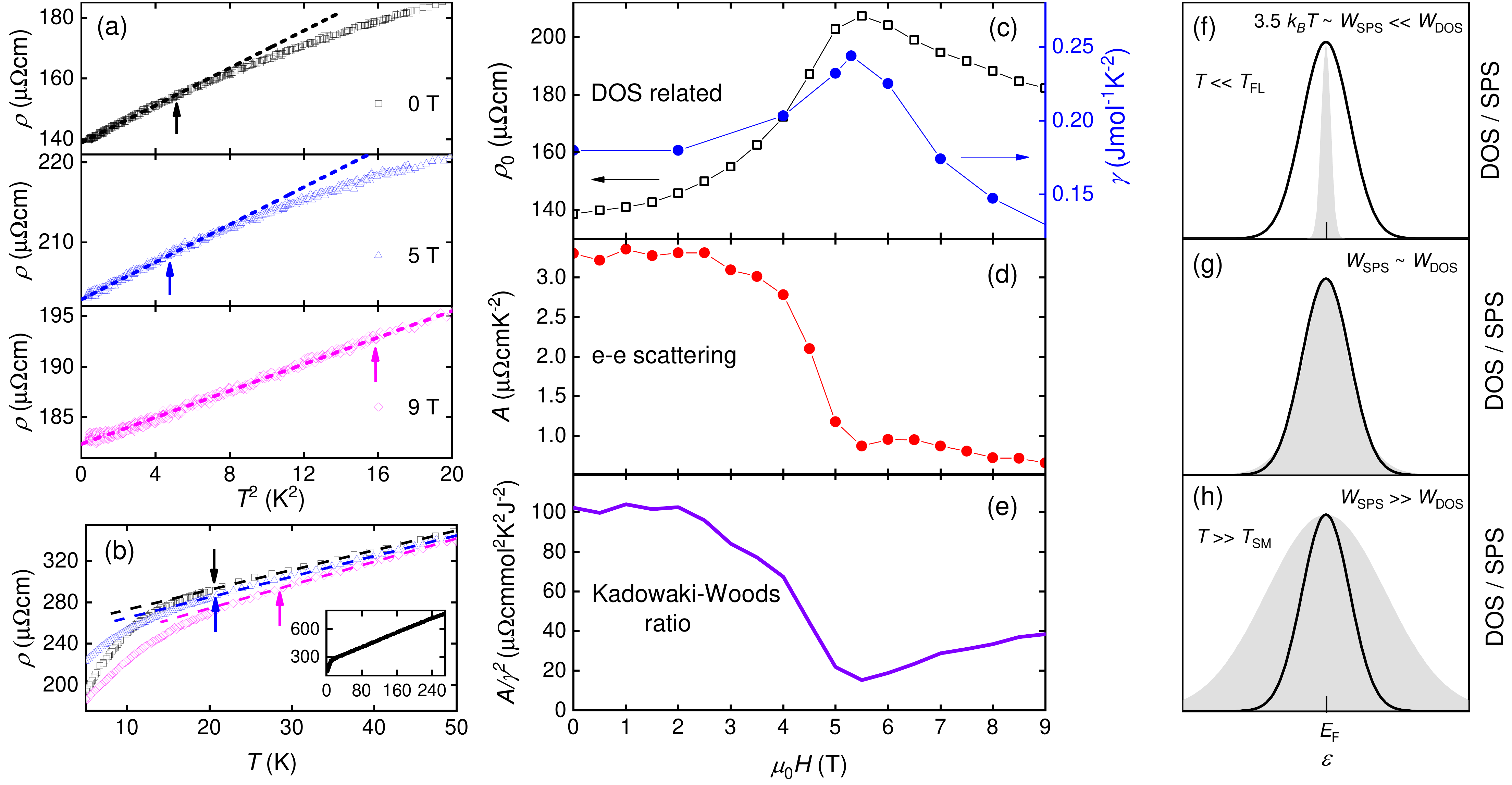}
 	\end{center}
\caption{\textbf{Kadowaki-Woods ratio across the metamagnetic transition in Ca$_{1.8}$Sr$_{0.2}$RuO$_4$.} (a) Resistivity $\rho$ plotted versus $T^{2}$ for selected fields. Dashed lines are linear fits to the low-temperature limit.
Arrows indicate the temperature scale $T${$\rm_{FL}$} above which the resistivity deviates from $T^2$ behavior. The temperature-independent term $\rho_0$ and coefficient of the $T^{2}$ term $A$ are obtained from the intercept and the slope of the linear fits, respectively. (b) Resistivity $\rho$ at higher temperatures plotted as a function of $T$ for selected fields. The dashed lines are linear fits to the high-temperature end. Arrows mark the $T${$\rm_{SM}$} scale below which resistivity deviates from a $T$-linear dependence. The inset shows the zero-field curve up to high temperature, where the ordinate is $\rho$ in the unit of $\mu\Omega$cm and the abscissa is $T$ in K. (c,d,e) Magnetic field dependence of $\rho_0$, $A$, $\gamma$, and the Kadowaki-Woods ratio $A/\gamma^2$. The Sommerfeld coefficient $\gamma$ is extracted from Ref.~\cite{BaierPhysB2006}. The error bars are smaller than the symbols. (f,g,h) Schematics comparing a density-of-states (DOS) peak profile (black line) with the scattering phase space (SPS, represented by the gray shaded area) for increasing temperatures. The scattering phase space is defined by $f_T(\epsilon)[1-f_T(\epsilon)]$, where $f_T(\epsilon)$ is the Fermi-Dirac distribution for temperature $T$ and energy $\epsilon$ measured from the Fermi level. Both profiles are centred around the Fermi level $E${$\rm_{F}$}, although the DOS peak position in a real material is tunable by $e.g.$, magnetic field or pressure. As discussed in the text, the DOS peak width depends on dimensionality and disorder.}
\end{figure*}

\noindent \textbf{Comparison of metamagnetic transitions:} Although the metamagnetic transition has been 
well established in Ca$_{1.8}$Sr$_{0.2}$RuO$_4$, its impact on magnetotransport has not been 
addressed by previous studies~\cite{NakatsujiPRL00,NakatsujiPRL2003,BalicasPRL2005} (see Supplementary Note 2). Our results demonstrate a direct connection between the metamagnetic transition and transport properties. As such, Ca$_{1.8}$Sr$_{0.2}$RuO$_4$ can now be directly compared to other metamagnetic systems. As shown in Table~1, Ca$_{1.8}$Sr$_{0.2}$RuO$_4$, CeRu$_2$Si$_2$ and Sr$_3$Ru$_2$O$_7$ all display a peak in $\rho_0$ and the Sommerfeld coefficient $\gamma$ across the metamagnetic transition. Both $\rho_0$ and $\gamma$ are proportional to the DOS at the Fermi level. Therefore, these compounds share a field-induced traversal of a DOS peak through the Fermi level. The DOS peak in Ca$_{1.8}$Sr$_{0.2}$RuO$_4$ is likely associated with a VHS~\cite{fang_magnetic_2001,fang_orbital-dependent_2004,LiebschPRL2007}. Interestingly, the inelastic electron-electron scattering process varies dramatically across these compounds. Non-Fermi liquid behavior is reported down to the lowest measured temperatures in Sr$_3$Ru$_2$O$_7$ at $H_m$. As in CeRu$_2$Si$_2$~\cite{DaouPRL2006}, we report Fermi liquid behavior across $H_m$ in Ca$_{1.8}$Sr$_{0.2}$RuO$_4$. However, in CeRu$_2$Si$_2$ the scattering coefficient $A$ peaks together with the Sommerfeld coefficient, whereas in Ca$_{1.8}$Sr$_{0.2}$RuO$_4$, $A$ undergoes a step-like drop across $H_m$. In the following, we discuss the Fermi liquid versus non-Fermi liquid aspect before turning to the unusual behavior of the 
KWR in Ca$_{1.8}$Sr$_{0.2}$RuO$_4$.  \\[2mm] 

\begin{table*}
\vspace{3mm}
\begin{center}
\begin{ruledtabular}
\begin{tabular}{cccccccc}
Compound & Tuning & Critical value & $\rho_0$ peak & $\gamma$ peak& $A$ peak&  FL & Reference  \\ \hline
Sr$_2$RuO$_4$ & Uniaxial strain  & $\epsilon=0.5$\% & Yes & Yes & Yes & No&~\cite{BarberPRL2018_full,herman_deviation_2019,li_high_2019}\\
Sr$_3$Ru$_2$O$_7$ &Magnetic field  & $H=7.8$~T & Yes & Yes & Yes & No&~\cite{GrigeraScience2001,rost_thermodynamics_2011,RostScience2011,PerryPRL2001,TokiwaPRL2016} \\
Ca$_{1.8}$Sr$_{0.2}$RuO$_4$ & Magnetic field & $H$ =  5.5~T & Yes &Yes & No &  Yes & This work, ~\cite{BaierPhysB2006}\\
CeRu$_2$Si$_2$ & Magnetic field & $H$ = 8.0~T & Yes & Yes & Yes & Yes &~\cite{DaouPRL2006,PfauPRB2012,BoukahilPRB2014,AokiJMMM1998} \\
CeTiGe & Magnetic field & $H$ = 12~T & Yes &  - &No & Yes&~\cite{DeppePRB2012} \\
\end{tabular}
\end{ruledtabular}

\caption{
\textbf{Fermi liquid behaviors as the density-of-states peak and the Fermi level are tuned to match.} For each compound the tuning parameter (uniaxial strain $\epsilon$ or magnetic field $H$) and the associated critical values are indicated. The behavior (peak or no peak) across the critical tuning of the temperature-independent term in resistivity $\rho_0$, Sommerfeld coefficient $\gamma$ and electron-electron scattering coefficient $A$ (see text) is indicated. Finally, 
the observed resistivity behavior (Fermi liquid or non-Fermi liquid) at the critical tuning and lowest measured temperature is given.}

\label{tab:tab1}	
\end{center}
\end{table*}

\section*{Discussion}
\noindent In strongly correlated electron systems, $\rho$ is generally dominated by impurity and electron-electron scattering at low temperatures. States contributing to the transport properties lie within the scattering phase space defined by $f_T(\epsilon)[1-f_T(\epsilon)]$, where $f_T(\epsilon)$ is the Fermi-Dirac distribution for temperature $T$ and energy $\epsilon$ measured from the Fermi level. For an electronic structure with a peak in the DOS close to or at the Fermi level, the phase-space energy scale, with a full-width half maximum $W${$\rm_{SPS}$} $\sim3.5k_B T$, can be compared with that of the DOS peak $W${$\rm_{DOS}$}. In the low-temperature limit $T \lesssim$ $T${$\rm_{FL}$} $\sim \kappa$ $W${$\rm_{DOS}$} $/(3.5k_B)$ with $\kappa\ll 1$, Fermi liquid behavior ($\rho \sim T^2$) is anticipated, since the DOS is almost flat within the scattering phase space. By contrast, for $T \gtrsim$ $T${$\rm_{SM}$} $\sim \beta$ $W${$\rm_{DOS}$} $/(3.5k_B)$ 
with $\beta\sim 1$, strange metal behavior, such as $\rho\sim T$, $\sim T^{3/2}$, or $\sim T^2$log$T$, is expected, once the DOS peak is fully covered by the scattering phase space~\cite{mousatov_theory_2020,bruin_similarity_2013,zaanen_why_2004,hartnoll_theory_2015,buhmann_unconventional_2013,hlubina_effect_1996}. These two limits, together with the intermediate region $T${$\rm_{FL}$} $< T <$ $T${$\rm_{SM}$} are schematically shown in Figs. 2(f)-2(h). 

Whereas the scattering phase space $W${$\rm_{SPS}$} is set by temperature, $W${$\rm_{DOS}$} is controlled by dimensionality and disorder. Utilizing $\rho_{ab}/\rho_c$ and $\rho_0$ as effective gauges for the dimensionality and disorder, respectively, we plot different systems with large DOS at the Fermi level in a dimensionality-disorder-temperature diagram (Fig. 3). For clean two-dimensional systems, such as Sr$_2$RuO$_4$ and Sr$_3$Ru$_2$O$_7$, the sharp DOS peak (small $W${$\rm_{DOS}$}) makes it difficult to experimentally access the temperature scales $T${$\rm_{FL}$} and $T${$\rm_{SM}$}. In both systems, when the Fermi level and VHS are tuned to match, strange metal behavior is observed down to lowest temperatures before being cut off by instabilities (superconductivity and spin density wave order)~\cite{BarberPRL2018_full,GrigeraScience2001}. In clean three-dimensional systems, a larger $T${$\rm_{FL}$} is expected, and indeed Fermi liquid behavior was found across $H_m$ in CeRu$_2$Si$_2$~\cite{DaouPRL2006}. To our knowledge,
in the two-dimensional dirty limit, Fermi liquid properties have not been explored/discussed in the context of a van Hove singularity. Notably, this limit is represented by Ca$_{1.8}$Sr$_{0.2}$RuO$_4$, where $T${$\rm_{FL}$} $\sim 2$~K [Figs. 1(c) and 2(a)] and $T${$\rm_{SM}$} $\sim 20$~K [Fig. 2(b)] are identified. Angle-resolved photoemission (ARPES) suggests that $W${$\rm_{DOS}$} $\sim 20$~meV~\cite{SutterPRB2019} stemming from disorder and electron correlations and hence, we extract reasonable values for $\kappa\sim 0.03$ and $\beta \sim 0.3 $. These values of $\kappa$ and $\beta$ are weakly material dependent as they stem from the ratio of the widths of the DOS and scattering phase space. Hence this information can be applied to, for example, the pseudogap problem~\cite{wu_pseudogap_2018,doiron-leyraud_pseudogap_2017,buhmann_numerical_2013} found in La-based cuprates. Assuming  $\beta\approx 0.3$ for La$_{1.36}$Nd$_{0.4}$Sr$_{0.24}$CuO$_4$, where $W${$\rm_{DOS}$} $\sim 15$~meV~\cite{matt_electron_2015}, yields $T${$\rm_{FL}$} $\sim 15$~K. However, since $\rho\sim T$ [$C/T\sim \log(1/T)$] is observed down to 1 (0.5)~K~\cite{daou_linear_2009,michon_wiedemann-franz_2018,legros_universal_2019,michon_thermodynamic_2019}, we conclude that quantum criticality must be taken into account. Our results thus have direct implications for the interpretation of the strange metal properties in cuprates.

\begin{figure}
\begin{center}
 		\includegraphics[width=0.48\textwidth]{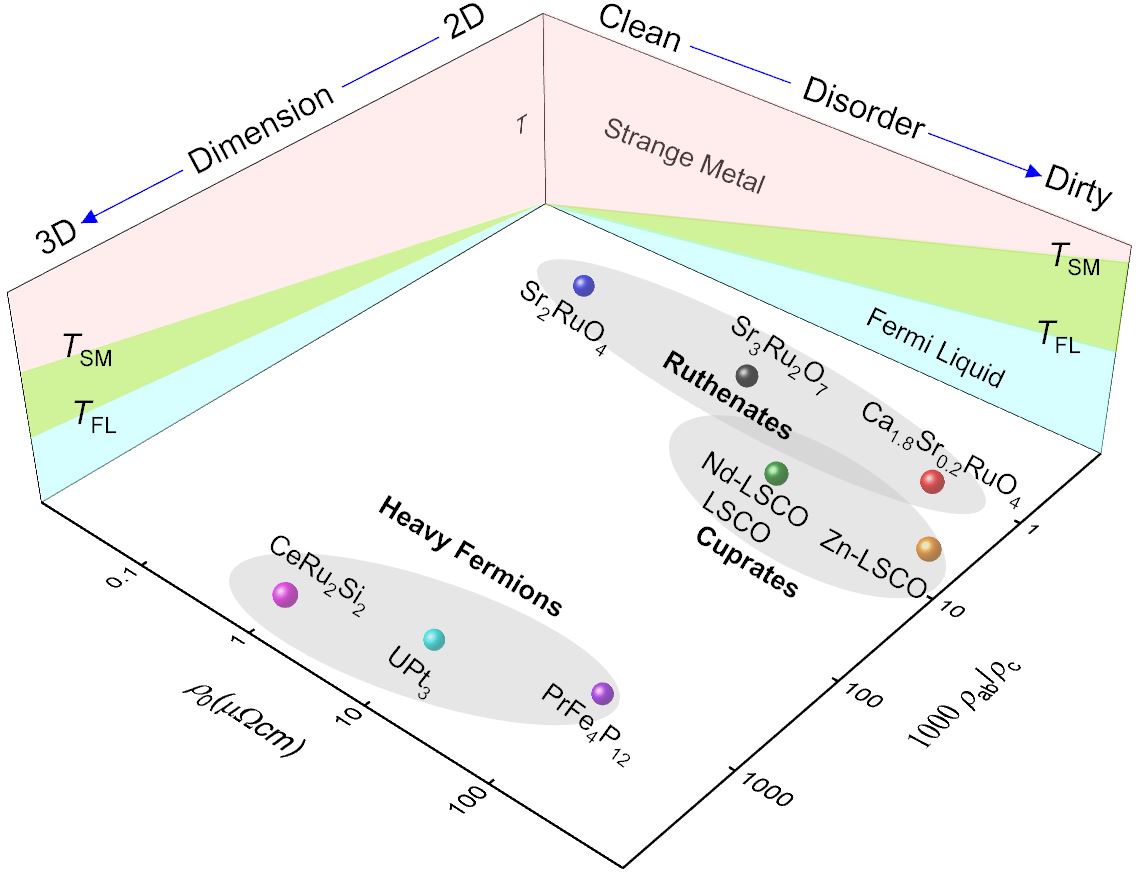}
 	\end{center}
\centering

\caption{\textbf{Schematics of Fermi liquid properties versus disorder and dimensionality. }
High density of states (DOS) systems~\cite{BarberPRL2018_full,tamai_fermi_2008,SutterPRB2019,wu_pseudogap_2018,doiron-leyraud_pseudogap_2017,DaouPRL2006,matsuda_specific-heat_2000,muller_specific_1989} plotted as a function of 
dimensionality and disorder gauged, respectively, by $\rho_{ab}/\rho_c$ (in-plane resistivity over out-of-plane resistivity) and $\rho_0$ (from Refs.~\cite{kikugawa_band-selective_2004,BarberPRL2018_full,rost_thermodynamics_2011,liu_electrical_2001,nakatsuji_switching_2000,lapierre_resistivity_1992,nakamura_anisotropic_1993,daou_linear_2009,michon_wiedemann-franz_2018,momono_low-temperature_1994,kim_indications_2000,de_visser_resistivity_1984,aoki_characterization_2012}). The third axis labeled $T$ refers to temperature. For all systems the DOS are tuned to the Fermi level by tuning parameters such as magnetic field or uniaxial pressure, and the values of $\rho_{ab}/\rho_c$ and $\rho_0$ are taken at these critical tuning parameter whenever possible. For the cuprates the following abbreviations are used: LSCO: La$_{1.8}$Sr$_{0.2}$CuO$_4$, Nd-LSCO: La$_{1.36}$Nd$_{0.4}$Sr$_{0.24}$CuO$_4$, Zn-LSCO: La$_{1.82}$Sr$_{0.18}$Cu$_{0.96}$Zn$_{0.04}$O$_4$. 
The vertical thermal axis indicates the two temperature scales $T${$\rm_{FL}$} and $T${$\rm_{SM}$} expected  within a van Hove singularity scenario. $T${$\rm_{FL}$} is the Fermi liquid cutoff temperature and above $T${$\rm_{SM}$} strange metal behavior dominates. 2D and 3D denote two-dimensional and three-dimensional systems, respectively.}

\end{figure}

The evolution of the KWR across the metamagnetic transition in Ca$_{1.8}$Sr$_{0.2}$RuO$_4$ is rather unusual. In the simplest case, the ratio $A/\gamma^2$ is invariant to electron correlations~\cite{rice_electron-electron_1968,kadowaki_universal_1986,hussey_non-generality_2005,jacko_unified_2009}. This implies that both $A$ and $\gamma^2$ are expected to increase with enhanced electron interaction. In practice, even in systems where $A/\gamma^2$ is not constant, $A$ and $\gamma^2$ still correlate positively. For example, a modified relation $A \sim \Delta\gamma$ holds in Sr$_3$Ru$_2$O$_7$, where $\Delta\gamma$ is the enhancement of $\gamma$ approaching $H_m$~\cite{mousatov_theory_2020}. In YbRh$_2$(Si$_{0.95}$Ge$_{0.05}$)$_2$ with a `local' QCP, $A$ and $\gamma$ both increase upon approaching the QCP, although the KWR shows a weak field dependence~\cite{custers_break-up_2003}. These are all in stark contrast to  Ca$_{1.8}$Sr$_{0.2}$RuO$_4$ where $A$ and $\gamma^2$ anti-correlate on approaching the metamagnetic transition on the low-field side. A factor-of-five variation [Fig. 2(e)] of the KWR is the consequence of this decoupling of $A$ and $\gamma^2$. 
We stress that the bare band structure is not expected to change significantly by the application of magnetic field and hence is not the source~\cite{hussey_non-generality_2005,jacko_unified_2009} for the strong variation of the KWR (see Supplementary Note 3 and Supplementary Figure 2). Worth noticing is also that elastic scattering -- probed by $\rho_0$ -- is  linked to the DOS at the Fermi energy. The field evolution of the KWR is therefore also reflected in the magnetic-field dependencies of $A$ and $\rho_0$ [Figs. 2(c) and 2(d)].
A similar decoupling of $A$ and $\rho_0$ has also been reported in the multiband heavy fermion system CeTiGe~\cite{DeppePRB2012}.

Although Ca$_{1.8}$Sr$_{0.2}$RuO$_4$ is a multiband system, we first resort to a (single-band) Boltzmann transport approach to gain qualitative insight into the KWR (see Supplementary Note 4). Within this framework, inelastic electron scattering is more sensitive than elastic scattering to the detailed relation between the DOS profile and the scattering phase space. This is most significant in systems with a DOS peak around the Fermi level, as is the case here. 

While a single-band model is certainly too simplistic, the additional complexities of multiband physics are not straightforward. In layered ruthenates, the multiband structure stems from the $d_{xy}$, $d_{xz}$ and $d_{yz}$ orbitals that produce a VHS in close vicinity to the Fermi level. This DOS peak is the most likely source of the metamagnetic transition. One would thus expect that a change of Fermi surface topology across the transition is a possible cause for the drop in $A$ and the resulting strong violation of the KWR. The multiorbital nature of the bands, however, suggests scattering between all bands, such that the diverging DOS should influence transport in all bands. The recent report of orbital-selective breakdown of Fermi-liquid behavior~\cite{SutterPRB2019}, on the other hand, implies a decoupling of the bands allowing for the above scenario of a step-like behavior of $A$. Note, finally, that momentum-dependent interactions, potentially stemming from the multi-orbital structure, can produce a momentum-dependent self-energy~\cite{mimick-Buhman_thesis,byczuk_kinks_2007,lohneysen_fermi-liquid_2007,stewart_non-fermi-liquid_2001,gull_momentum-space_2010,tanatar_anisotropic_2007,smith_apparent_2008,chang_anisotropic_2013}, which provides another source for the unusual behavior of the KWR~\cite{jacko_unified_2009}.

Metamagnetic transitions are found in materials spanning from correlated oxides to heavy fermion compounds. The underlying mechanism might not be identical across all compounds and hence comparative studies are of great interest. We performed a comprehensive study of the metamagnetic transition of Ca$_{1.8}$Sr$_{0.2}$RuO$_4$. Presence of a tunable van Hove singularity and disorder provides an explanation for the observed temperature scales associated with the Fermi liquid and strange metal properties. Previous studies suggested quantum critical scaling around the metamagnetic transition may be smeared out by disorder~\cite{SteffensPRL2007,baier_magnetoelastic_2007}. This is likely the reason why Fermi liquid behavior survives across the metamagnetic transition in Ca$_{1.8}$Sr$_{0.2}$RuO$_4$ but breaks down in the clean limit represented by Sr$_3$Ru$_2$O$_7$. Alternatively, if only part of the quasiparticles participate in the mass divergence upon approaching a putative quantum critical point at the metamagnetic transition, they can get short-circuited by the remaining quasiparticles. This would reflect on $A$ but not $\gamma$. In this scenario, persisting  Fermi liquid behavior and a varying KWR are expected. Our study demonstrates that electronic properties across a van Hove singularity induced metamagnetic transition is strongly influenced by the degree of disorder. In the highly disordered limit, we observed an unusual strong violation of the Kadowaki-Woods ratio. 

~\\

\section*{Methods}
\noindent Single crystals of Ca$_{1.8}$Sr$_{0.2}$RuO$_4$ were grown by the flux-feeding floating-zone technique~\cite{NAKATSUJI200126,FUKAZAWA2000613}. Our experimental results were reproduced on several crystals that were cut and polished into a rectangular shape, with the largest natural plane being the $ab$ plane. Magnetic fields $\mu_0 H$ ($\mu_0$ being the vacuum permeability) up to 9 T were applied along the $c$ axis and silver paste electrical contacts were made on the $ab$ plane. Resistivity measurements were performed in a physical property measurement system (PPMS, Quantum Design) with a Helium-3 option.\\

\section*{Data availability} 
\noindent The data that support the findings of this study are available from the corresponding author upon reasonable request. \\

\bibliographystyle{nature4}

~\\

\section*{Acknowledgements}
\noindent We thank Louis Taillefer and Beno\^it Fauqu\'e for helpful discussions. Y.X., F.H., and J.C. were financially supported by the Swiss National Science Foundation (SNSF) (Grant No. PP00P2\_179097 and 184739). L.D. was partially funded through Excellence Scholarship by the Swiss Government. \\

\section*{Author contributions}
\noindent V.G., R.F., and A.V. grew and prepared the Ca$_{1.8}$Sr$_{0.2}$RuO$_4$ single crystals. Y.X. and D.D. carried out the magnetoresistance experiments with assistance from L.D., J.V., S.G., J.S., M.G., A.S., X.Z., and S.L. Y.X. analysed the experimental data.
Y.X., F.H., M.H.F., and J.C. wrote the manuscript with inputs from all authors.\\

\section*{Competing interests}
\noindent The authors declare no competing interests.\\

\end{document}